\documentclass[aps,pra,amssymb,twocolumn,superscriptaddress,showpacs]{revtex4}

\usepackage{graphicx}
\usepackage{dcolumn}
\usepackage{bm}
\usepackage{color}

\begin{document}
\voffset1cm

\newcommand{\beq}{\begin{equation}}
\newcommand{\eeq}{\end{equation}}
\newcommand{\barr}{\begin{eqnarray}}
\newcommand{\earr}{\end{eqnarray}}

\newcommand{\REV}[1]{\textbf{\color{red}#1}}
\newcommand{\BLUE}[1]{\textbf{\color{blue}#1}}
\newcommand{\GREEN}[1]{\textbf{\color{green}#1}}

\newcommand{\andy}[1]{ }

\newcommand{\bmsub}[1]{\mbox{\boldmath\scriptsize $#1$}}

\def\R{\mathbb{R}}

\def\bra#1{\langle #1 |}
\def\ket#1{| #1 \rangle}
\def\sinc{\mathop{\text{sinc}}\nolimits}
\def\cV{\mathcal{V}}
\def\cH{\mathcal{H}}
\def\cT{\mathcal{T}}
\renewcommand{\Re}{\mathop{\text{Re}}\nolimits}
\newcommand{\tr}{\mathop{\text{Tr}}\nolimits}

\title{Generalized tomographic maps }

\author{M. Asorey}
\affiliation{Departamento de F\'\i sica Te\'orica, Facultad de
Ciencias, Universidad de Zaragoza, 50009 Zaragoza, Spain}
\author{P. Facchi}
\affiliation{Dipartimento di Matematica, Universit\`a di Bari,
        I-70125  Bari, Italy}
\affiliation{INFN, Sezione di Bari, I-70126 Bari, Italy}
\author{V.I. Man'ko}
\affiliation{P.N. Lebedev Physical Institute, Leninskii Prospect
53, Moscow 119991, Russia}
\author{G. Marmo} \affiliation{Dipartimento di Scienze Fisiche,
Universit\`a di Napoli ``Federico II", I-80126  Napoli, Italy}
\affiliation{INFN, Sezione di Napoli, I-80126  Napoli, Italy}
\author{S. Pascazio} \affiliation{Dipartimento di Fisica,
Universit\`a di Bari,
        I-70126  Bari, Italy}
\affiliation{INFN, Sezione di Bari, I-70126 Bari, Italy}
\author{E.C.G. Sudarshan} \affiliation{Department of Physics,
University of Texas, Austin, Texas 78712, USA}

\date{\today}

\begin{abstract}
We introduce several possible generalizations of tomography for
quadratic surfaces. We analyze different types of elliptic,
hyperbolic and hybrid tomograms. In all cases it is possible to
consistently define the inverse tomographic map. We find two
different ways of introducing tomographic sections. The first method
operates by deformations of the standard Radon transform. The second
method proceeds by shifting a given quadric pattern. The most
general tomographic transformation can be defined in terms of
marginals over surfaces generated by deformations of complete
families of hyperplanes or quadrics. We discuss practical and
conceptual perspectives and possible applications.
\end{abstract}


\maketitle

\section{Introduction}
\label{sec-introd}

Most of classical  applications of tomography are based on light
propagation along optic rays (implicitly assumed to be straight
lines). Standard Radon transform theory guarantees that a
measurement of the absorption of light beams travelling in
dielectric media in straight lines allows the complete
reconstruction of the matter density of these media. Indeed, the
original Radon transform \cite{Rad1917} maps functions of two
variables in the plane onto functions of one real variable on a line
and one variable on a circle. The crucial property is that the
transform is invertible and continuous \cite{John, Strichartz}.

There exist several generalizations of the Radon transform. See,
e.g., \cite{Helgason} and \cite{Gelf}. Further generalizations can
be motivated by physical observations: for instance, if the function
on the plane is a probability density, its Radon component is  a
family of probability densities of one random variable on the line,
parameterized by a variable living on a circle \cite{Olga97}. A
tomographic approach in a similar framework was applied to a free
classical particle moving on a circle \cite{tomogram}, where the
phase space is a two dimensional cylinder.

In quantum mechanics the Radon transform of the Wigner function
\cite{Wig32} was considered in the tomographic approach to the study
of quantum states \cite{Ber-Ber,Vog-Ris} and experimentally
realized with different particles and in diverse situations
\cite{SBRF93,torino,konst}. Other experiments have been proposed
\cite{reconstruct06} and the whole field is in continuing
evolution, also in view of its relevance in genuine quantum
mechanical problems and quantum information related topics. Good
reviews on recent tomographic applications can be found in
Ref.~\cite{Jardabook}, with emphasis on maximum likelihood estimations
\cite{theory}, that enable one to extract the maximum reliable
information from the available data.

A further development, extending the analysis to incorporate more
general symplectic transforms, was presented in
\cite{Mancini95} and the mathematical mechanism at the basis of the
mapping of true density states onto tomographic probabilities was
elucidated in \cite{Marmoopen}. There is an interesting relation
between the Radon map of Wigner functions and the formalism of star
product quantization \cite{MarmoPhysScr,Olga97}: symplectic
tomograms are indeed the Radon components of the Wigner function and
this enables one to define a procedure aimed at determining the
marginal probability densities along straight lines in phase space.
The knowledge of all these marginals makes possible the
reconstruction of the Wigner function in the quantum case and of the
probability density in the classical case.

The generalization of tomographic maps to curved surfaces opens new
perspectives in the applications of tomography both to quantum and
classical systems. Some attempts to study marginals along curves
other than straight lines were introduced  in Ref.\
\cite{ManMenPhysD}. Very recently, optical ``accelerating" Airy
beams were observed \cite{airy}: these beams could be used to
perform a tomographic map over parabolas in phase space. A
generalization of tomography to this kind of applications requires a
generalization of the Radon transform.

The aim of this article is to study generalizations of the Radon
transform to multidimensional phase spaces and to marginals along
curves or surfaces. Most of the generalizations of the Radon transform
proceed by considering geodesic submanifolds of a given Riemannian
manifold. We develop here a different approach, that can be
applied to the Radon components of the probability densities of
classical particles in phase space, and construct the
corresponding tomographic maps.

This article is organized as follows. In Sec.\
\ref{sec-tomogplane} we review the standard tomographic
application of the Radon transform on the plane. In Sec.\
\ref{sec-hyperplane} we consider the generalization to arbitrary
dimensions. A deformation of the Radon transform with applications
to elliptic and hyperbolic problems is presented in Sec.\
\ref{sec-submanifolds}. In Sec.\ \ref{sec-quadrics}
we introduce a new type of transform involving hyperbolic,
elliptic and parabolic quadrics. The transform is defined by
translations of a basic pattern. Finally in Sec.\
\ref{sec-concl} we discuss the relevance of our results for
future applications.

\section{Tomography on the plane}
\label{sec-tomogplane}

Let us consider a function $f(q,p)$ on the phase space $(q,p) \in
\mathbb{R}^2$ of a particle moving on the line $q \in \mathbb{R}$.
The Radon transform, in its original formulation, solves the
following problem: reconstruct a function of two variables, say
$f(p,q)$, from  its integrals over arbitrary lines.

\begin{figure}[t]
\begin{center}
\includegraphics[width=8cm]{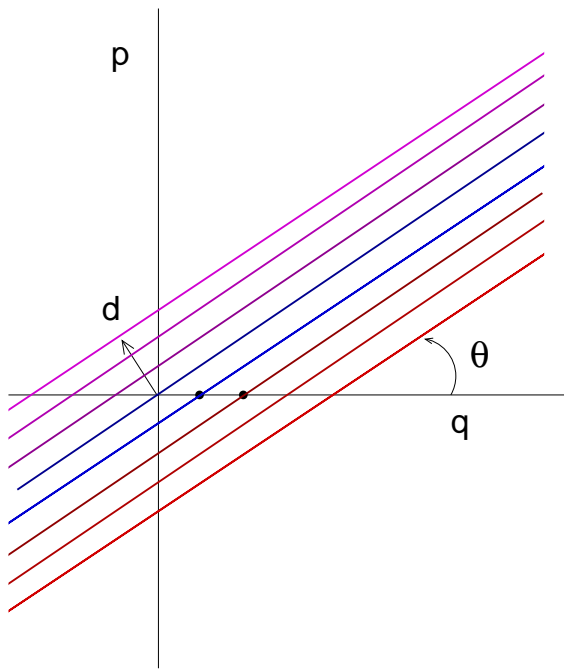}
\end{center}
\caption{Tomography on the plane; $(q,p) \in
\mathbb{R}^2, d \in \mathbb{R}, \theta \in \mathbb{S}$ (unit circle).}
\label{fig:plane}
\end{figure}

In the $(q,p)$ plane, a line is given by the equation
\begin{equation}
X -\mu q - \nu p = 0,
\end{equation}
with $(\mu,\nu)\neq(0,0)$. Thus, the family of lines has the
manifold structure $\mathbb{R}\times\mathbb{S}$, with $\mathbb{S}$
the unit circle, $d=X/\sqrt{\mu^2+\nu^2}\in\mathbb{R}$ and
$\mu/\nu=\tan\theta$, $\theta\in\mathbb{S}$ (see Fig.\
\ref{fig:plane}).

It is possible to write the Radon transform in affine language
(tomographic map) \cite{Rad1917,Gelf}
\begin{eqnarray}
\omega_f(X,\mu,\nu) &=& \left\langle \delta(X-\mu q-\nu
p)\right\rangle \nonumber\\
& =& \int_{\mathbb{R}^2} f(q,p) \delta(X-\mu q-\nu p)  dqdp ,
\label{eq:radondef}
\end{eqnarray}
where $\delta$ is the Dirac function and the parameters $X, \mu,
\nu \in \mathbb{R}$.

The inverse transform of (\ref{eq:radondef}) reads
\begin{eqnarray}
f(q,p) &=&  \int_{\mathbb{R}^3} \omega_f(X,\mu,\nu) e^{i(X-\mu
q-\nu p)} \frac{dXd\mu d\nu}{(2\pi)^2} . \quad \label{eq:invgen}
\end{eqnarray}
The homogeneity of $\omega_f(X,\mu,\nu)$
\begin{eqnarray}
\omega_f(\lambda X,\lambda \mu,\lambda \nu) =
\frac{1}{|\lambda|}\omega_f(X,\mu,\nu),
\label{eq:homog}
\end{eqnarray}
$\forall \lambda\in \mathbb{R}$, $\lambda\neq0$, is a direct
consequence of (\ref{eq:radondef}). If the function $f(q,p)$ is a
probability density function on the phase space of a classical
particle, i.e.
\begin{eqnarray}
f(q,p) \geq 0, \quad \int_{\mathbb{R}^2} f(q,p) dqdp =1,
\label{eq:denscond}
\end{eqnarray}
{the function $\omega_f(X,\mu,\nu)$   } is also nonnegative and is
called symplectic tomogram or Radon transform of the distribution
function $f(q,p)$ (in analogy to the Fourier transform of a
function).  The Radon transform contains the same information on the
state of the particle evolving on the phase space as the initial
distribution function. In summary, the tomograms
\begin{eqnarray}
\omega_f(X,\mu,\nu) \geq 0, \quad \int_{\mathbb{R}}
\omega_f(X,\mu,\nu) dX =1, \quad \forall \mu, \nu,
\label{eq:normcond}
\end{eqnarray}
form a family of density functions that depends on the two real parameters $\mu$ and $\nu$.

\section{Tomograms on  hyperplanes}
\label{sec-hyperplane}

The above construction can be generalized to higher dimensional
spaces in a straightforward way. Let us consider a function $f(q)$
on the $n$-dimensional space $q
\in \mathbb{R}^n$. Is it possible to reconstruct the function $f$ from its integrals
over arbitrary $(n-1)$-dimensional linear submanifolds? The answer
to this question is positive and provides a generalization of the
original Radon transform.

A generic hyperplane is given by the equation
\begin{equation}
X -\mu \cdot q = 0,
\label{eq:hypplan}
\end{equation}
with $X\in\mathbb{R}$ and $\mu\in\mathbb{R}^n$. Due to homogeneity,
this family of hyperplanes  is an $n$-dimensional manifold
diffeomorphic to $\mathbb{R}\times
\mathbb{S}^{n-1}$, because any hyperplane can be characterized by its unit
normal vector $\mu/|\mu|$ and its distance to the origin
${|X|}/{|\mu|}$. Note that this manifold is
not diffeomorphic to $\mathbb{R}^n$ because the sphere $\mathbb{S}^{n-1}$ is compact.

The Radon transform is given by
\begin{eqnarray}
\omega_f(X,\mu) &=& \left\langle \delta(X-\mu \cdot q)\right\rangle \nonumber\\
& =& \int_{\mathbb{R}^n} f(q) \delta(X-\mu\cdot q)  d^nq .
\label{eq:radondefn}
\end{eqnarray}
When $n=2$
Eq.\ (\ref{eq:radondef}) is recovered.

The inverse transform of
(\ref{eq:radondefn}) reads
\begin{eqnarray}
f(q) &=&  \int_{\mathbb{R}^{n+1}} \omega_f(X,\mu) e^{i(X-\mu\cdot
q)} \frac{dXd^n\mu}{(2\pi)^n} . \quad \label{eq:invgenn}
\end{eqnarray}
The homogeneity of $\omega_f(X,\mu)$
\begin{eqnarray}
\omega_f(\lambda X,\lambda \mu) =
\frac{1}{|\lambda|}\omega_f(X,\mu),
\label{eq:homog1}
\end{eqnarray}
$\forall \lambda\in \mathbb{R}$, $\lambda\neq0$,  is a direct
consequence of (\ref{eq:radondefn}). If the function $f(q)$ is a
probability density function on $\mathbb{R}^n$
\begin{eqnarray}
f(q) \geq 0, \quad \int_{\mathbb{R}^n} f(q) d^nq =1,
\label{eq:denscond1}
\end{eqnarray}
also the tomograms $\omega_f(X,\mu)$ are probability densities
\begin{eqnarray}
\omega_f(X,\mu) \geq 0, \quad \int_{\mathbb{R}} \omega_f(X,\mu) dX
=1, \quad \forall \mu\in\mathbb{R}^n
\label{eq:normcond1}
\end{eqnarray}
and the family of tomograms depends on the $n$ real parameters
$\mu$. In quantum mechanics such construction was applied to Wigner
functions providing a center of mass tomography
\cite{archi}.

\section{Tomograms on hypersurfaces}
\label{sec-submanifolds}

A simple  mechanism that allows nonlinear generalizations of the
Radon transform is the combination of the standard transform with a
diffeomorphism of the underlying $\mathbb{R}^n$ space. Let us
consider a function $f(q)$ on the $n$-dimensional space $q
\in \mathbb{R}^n$. The problem is to reconstruct $f$ from
 its integrals over an $n$-parameter family of submanifolds of codimension one.

\begin{figure}[t]
\begin{center}
\includegraphics[width=\columnwidth]{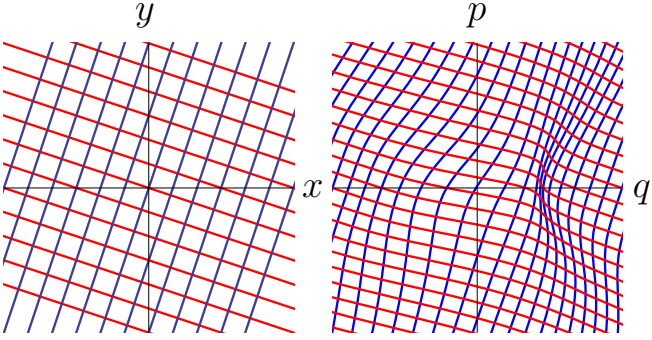}
\end{center}
\caption{Diffeomorphism of the plane: $(q,p)\in
\mathbb{R}^2 \to (x,y)=\varphi(q,p) \in \mathbb{R}^2$.}
\label{fig:diffeo}
\end{figure}

We can construct such a family by diffeomorphic deformations of
the hyperplanes (in the $x\in\mathbb{R}^n$ space)
\begin{equation}
X -\mu \cdot x = 0, \label{eq:planes}
\end{equation}
with $X\in\mathbb{R}$ and $\mu\in\mathbb{R}^n$. Let us
consider a diffeomorphism of $\mathbb{R}^n$
\begin{equation}
q\in\mathbb{R}^n\mapsto x=\varphi(q)\in\mathbb{R}^n.
\end{equation}
The hyperplanes (\ref{eq:planes}) are deformed by $\varphi$ into a
family of submanifolds (in the $q$ space)
\begin{equation}
X -\mu \cdot \varphi(q) = 0.
\end{equation}
The case $n=2$ is displayed in Fig.\ \ref{fig:diffeo}: $(q,p)\in
\mathbb{R}^2 \to (x,y)=\varphi(q,p) \in \mathbb{R}^2$.

Given a probability density $\tilde f (x)$ on the $x$ space, the
Radon transform can be rewritten as
\begin{eqnarray}
\!\!\!\!\!\!\!\!\!\!\!\!\omega_f(X,\mu) &=& \left\langle \delta(X-\mu \cdot x)\right\rangle \nonumber\\
&=& \int_{\mathbb{R}^n} \tilde f(x) \delta(X-\mu\cdot x)  d^nx
\nonumber\\
&=& \int_{\mathbb{R}^n} \tilde f(\varphi(q)) \delta(X-\mu\cdot
\varphi(q))J(q) d^nq ,
\label{eq:radondefns0}
\end{eqnarray}
where
\begin{equation}\label{eq:Jacobian}
J(q)=\left|\frac{\partial x_i}{\partial
q_j}\right|=\left|\frac{\partial \varphi_i(q)}{\partial
q_j}\right|
\end{equation}
is the Jacobian of the transformation.

Observe now that
\begin{equation}
\tilde f(x) d^n x = \tilde f(\varphi(q)) J(q) d^n q,
\end{equation}
whence $f (q) = \tilde f(\varphi(q)) J(q)$ is a probability
density. Therefore the tomograms are given by
\begin{eqnarray}
\omega_f(X,\mu) &=& \left\langle \delta(X-\mu \cdot \varphi(q))\right\rangle \nonumber\\
& =& \int_{\mathbb{R}^n} f(q) \delta(X-\mu\cdot \varphi(q))d^nq ,
\label{eq:radondefns}
\end{eqnarray}
with $X\in \mathbb{R}$ and $\mu \in \mathbb{R}^n$.

The inverse transform follows by (\ref{eq:invgenn}):
\begin{eqnarray}
f(q) &=& \tilde f(\varphi(q)) J(q) \nonumber\\
&=&  \int_{\mathbb{R}^{n+1}} \omega_f(X,\mu) J(q) e^{i(X-\mu\cdot
\varphi(q))} \frac{dXd^n\mu}{(2\pi)^n} , \quad
\label{eq:invgens}
\end{eqnarray}
with a modified kernel
\begin{equation}\label{eq:kernel}
\!\!\!\!\!K(q;X,\mu)\! =\! J(q) e^{i(X-\mu\cdot \varphi(q))}=
\left|\frac{\partial \varphi_i(q)}{\partial q_j}\right|
e^{i(X-\mu\cdot \varphi(q))}.
\end{equation}
Therefore, a probability density distribution on $\mathbb{R}^n$
\begin{eqnarray}
f(q) \geq 0, \quad \int_{\mathbb{R}^n} f(q) d^nq =1,
\label{eq:denscond2}
\end{eqnarray}
produces tomograms $\omega_f(X,\mu)$ that are probability
densities
\begin{eqnarray}
\omega_f(X,\mu) \geq 0, \quad \int_{\mathbb{R}} \omega_f(X,\mu) dX
=1, \quad \forall \mu\in\mathbb{R}^n.
\label{eq:normcond2}
\end{eqnarray}
The family of tomograms depends on the $n$ real parameters $\mu$. We
can now consider different applications of these deformed
generalizations of the Radon transform.

\subsection{Circles in the plane}
\label{sec-circles}

\begin{figure}[t]
\begin{center}
\includegraphics[width=8cm]{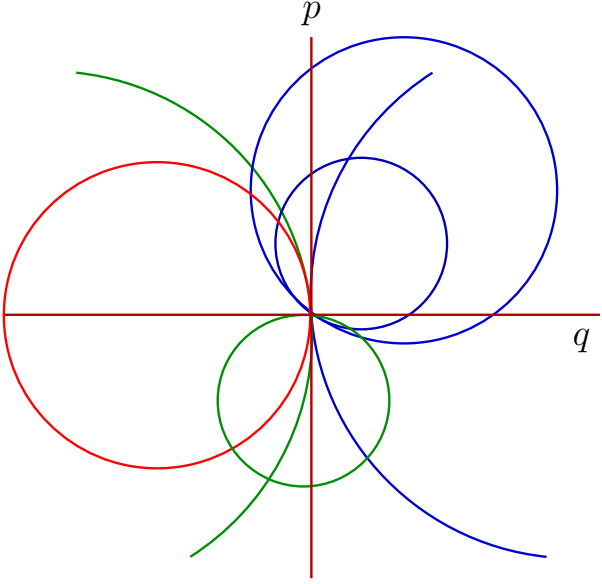}
\end{center}
\caption{Deformed circular tomography. All circles pass through the origin.}
\label{fig:circles}
\end{figure}

In the punctured $(x,y)$ plane without the origin $(0,0)$,  the conformal inversion
\begin{equation}\label{eq:conformal}
(x,y)=\varphi(q,p)=\left(\frac{q}{q^2+p^2},\frac{p}{q^2+p^2}\right),
\end{equation}
maps the family of  lines
\begin{equation}
X -\mu x - \nu y = 0
\end{equation}
into a family of circles
\begin{equation}
X (q^2+p^2) -\mu q  - \nu p = 0,
\label{eq:circles}
\end{equation}
centered at
\begin{equation}\label{eq:centers}
C=\left(\frac{\mu}{2X},\frac{\nu}{2X} \right)
\end{equation}
and passing through the origin (see Fig.\ \ref{fig:circles}). When
$X=0$ they degenerate into lines through the origin.

The Jacobian reads
\begin{equation}
J(q,p)=\left|\frac{\partial(x,y)}{\partial(q,p)}\right|=\frac{1}{(q^2+p^2)^2},
\end{equation}
whence the transformation is a diffeomorphism of the punctured plane
of functions $f \in L^1(\mathbb{R}^2)$. The  singularity of the
transformation at the origin $(0,0)$ is irrelevant for  tomographic
integral transformations because it only affects a zero measure set.

Equations (\ref{eq:radondefns})-(\ref{eq:invgens}) become
\begin{eqnarray}
& & \!\!\!\!\!\!\!\!\!\!\!\!\!\!\!\!\!\!\!\!\!\!\!\!\omega_f(X,\mu,\nu) = \left\langle
\delta\left(X-\frac{\mu q}{q^2+p^2} -\frac{\nu p}{q^2+p^2}\right)\right\rangle \nonumber\\
& & \!\!\!\!\!\!\!\!\!\!\!\!\!\!\!\!\!\!\!\!\!\!\!\!= \int_{\mathbb{R}^2} f(q,p) \delta\left(X-\frac{\mu q}{q^2+p^2} -\frac{\nu p}{q^2+p^2}\right) dq dp
\label{eq:radondefnc}
\end{eqnarray}
and
\begin{eqnarray}
f(q,p) = \int_{\mathbb{R}^3} \omega_f(X,\mu)
\frac{e^{i(X-\frac{\mu q}{q^2+p^2} -\frac{\nu
p}{q^2+p^2})}}{{(2\pi)^2}(q^2+p^2)^2}\,
{dX
d\mu d\nu} . \quad
\label{eq:invgenc}
\end{eqnarray}

\subsection{Hyperbolas in the plane}
\label{sec-hyperbolas}

\begin{figure}[t]
\begin{center}
\includegraphics[width=\columnwidth]{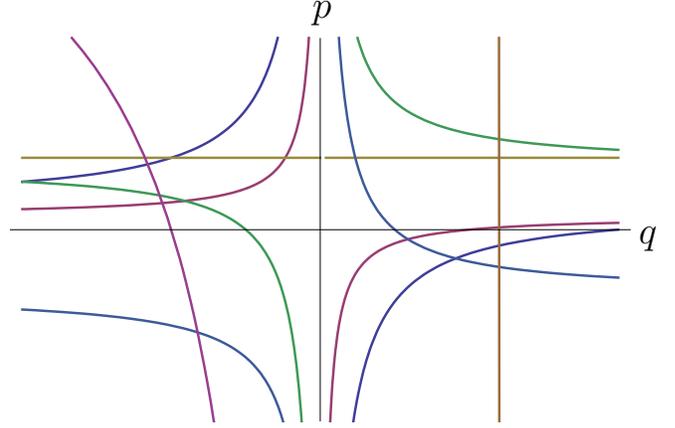}
\end{center}
\caption{Hyperbolic tomography. }
\label{fig:hyperb}
\end{figure}

In the  $(x,y)$ plane  the family of  lines
\begin{equation}
X -\mu x - \nu y = 0
\end{equation}
is mapped into a family of hyperbolas
\begin{equation}
X -\frac{\mu}{q}  - \nu p = 0,
\label{eq:hyperbola}
\end{equation}
with asymptotes
\begin{equation}
q= 0, \qquad p= \frac{X}{\nu} ,
\end{equation}
by the transformation
\begin{equation}\label{eq:inversion}
(x,y)=\varphi(q,p)=\left(\frac{1}{q},p\right).
\end{equation}
For $\mu>0$ the hyperbolas are in the second and fourth quadrants,
while for $\mu<0$ they are in the first and third quadrants (see
Fig.\ \ref{fig:hyperb}). When $\mu=0$ or $\nu=0$ they degenerate
into horizontal or vertical lines, respectively.

The Jacobian reads
\begin{equation}
J(q,p)=\left|\frac{\partial(x,y)}{\partial(q,p)}\right|=\frac{1}{q^2},
\end{equation}
whence the transformation is a diffeomorphism in the cut plane
without the axis $(0,y)$.

Equations (\ref{eq:radondefns})-(\ref{eq:invgens})become
\begin{eqnarray}
\!\!\omega_f(X,\mu,\nu) &=& \left\langle \delta\left(X-\frac{\mu}{q} -\nu p\right)\right\rangle \nonumber\\
& =& \int_{\mathbb{R}^2} f(q,p) \delta\left(X-\frac{\mu}{q} -\nu
p\right) dq dp
\label{eq:radondefnh}
\end{eqnarray}
and
\begin{eqnarray}
f(q,p) = \int_{\mathbb{R}^3} \omega_f(X,\mu) \frac{1}{q^2}
e^{i(X-\frac{\mu}{q}-\nu p)} \frac{dX d\mu d\nu}{(2\pi)^2} . \quad
\label{eq:invgenh}
\end{eqnarray}
The tomograms (\ref{eq:radondefns}), (\ref{eq:radondefnc}) and
(\ref{eq:radondefnh}) have the homogeneity property
(\ref{eq:homog1}).

\subsection{Hyperboloids in $\R^n$}
The generalization to higher dimensions of tomographic maps that can
be given in terms of quadratic expressions is straightforward.
Let us consider for example the following tomographic map
\begin{eqnarray}
\omega_f(X,\mu,\nu)\! =\!\! \int_{\mathbb{R}^{2n}}    \delta \left(X-\mu\cdot q-\nu (q,p) \right)
 f(q,p) {d^n q\ d^n p },
\nonumber\\
\label{eq:exf2}
\end{eqnarray}
where $p$ and $q$ are vectors in $\R^n$ and
\begin{equation}
\nu (q,p)=\sum_{j=1}^n \nu_j q_j  p_j.
\end{equation}
This map corresponds to
a deformation of the standard multidimensional Radon transform by
means of the following  diffeomorphism of $\mathbb{R}^{2n}\backslash\bigcup_{j}\{(q,p): q_j=0\}$
\beq
(q_i,p_j)\mapsto (x_i,y_j)=\left({q_i}, {q_j} p_j\right),
\eeq
whose Jacobian is
\beq
J(q,p)=\left|\frac{\partial(x,y)}{\partial(q,p)}\right|=\prod_{j=1}^n|
q_j | .
\eeq
The inverse map is given by
\begin{eqnarray}
 f(q,p)  &=&\int_{\mathbb{R}^{2n+1}} \frac{dX\, d^n \mu\, d^n\nu}{(2\pi)^{2n}}\,\omega_f(X,\mu,\nu)
\nonumber\\
& & \quad \times \prod_{j=1}^n| q_j |\,  {\rm e}^{i\left(X-\mu\cdot
q-\nu (q,p) \right)} .
\label{eq:inv2}
\end{eqnarray}
This corresponds to the higher dimensional generalization of the
Bertrand-Bertrand tomography \cite{Ber-Ber}.

Note that when $n=2$, by interchanging the role of $X$ and $-\mu$,
one recovers the same distribution of hyperbolas in the plane
analyzed in the previous subsection \ref{sec-hyperbolas}.

Although the above generalizations might be very useful for light
rays tomograms, all of them  involve integration over unbounded
submanifolds. One would like to generalize the Radon transform to
marginals defined over \emph{compact} submanifolds, that are bounded
on a compact domain around $(p,q)$. This case will be investigated
in the following section.

\section{Tomograms on quadrics}
\label{sec-quadrics}

Let us now look for a different generalization of tomograms.
We shall consider marginals
along compact quadrics. This can be achieved by shifting and scaling a given
quadric pattern
\begin{equation}
\label{eq:quartic}
X=(q-\mu, B (q-\mu)) ,
\end{equation}
where $B$ is a non-degenerate symmetric operator with respect to the
scalar product $(x,y)=x\cdot y$. A new generalization of the
tomographic map can be defined by
\begin{eqnarray}
 \omega_f(X,\mu ;B) = \int_{\mathbb{R}^n}  f(q)\,
 \delta \left(X-(q-\mu, B (q-\mu))\right) {d^n q}.
 \nonumber\\
\label{new}
\end{eqnarray}
Eq.~(\ref{new}) defines a completely
different type of transform, supported on the quadrics defined by
equation (\ref{eq:quartic}).
 It is easy to show that the inverse map is defined by
\begin{eqnarray}
 f(q)  &=&\displaystyle\frac{|\det B|}{\pi^n } \int_{\mathbb{R}^{n+1}} dX\, d^n \mu  \, \omega_f(X,\mu ;
 B)\nonumber\\
 & & \qquad \times\displaystyle{\rm e}^{i\left(X-(q-\mu, B (q-\mu))\right)}.
\label{eq:inv}
\end{eqnarray}
Indeed, by applying the definition of tomographic map (\ref{eq:inv})
\begin{eqnarray}
& &\!\!\!\!\!\!\!\!\!\! \displaystyle\frac{|\det B|}{\pi^n }\int dX\, d^n \mu\,  {\rm
e}^{i\left(X-(q-\mu, B (q-\mu))\right)} \omega_f(X,\mu ;
 B) \nonumber\\
&&= \displaystyle\frac{|\det B|}{\pi^n }\int dX\, d^n \mu\,  {\rm
e}^{i\left(X-(q-\mu, B (q-\mu))\right)}
 \nonumber\\
& & \times \int {d^n \xi\ } \delta\left(X-(\xi-\mu, B
(\xi-\mu))\right) f(\xi),
\end{eqnarray}
which after integration over $X$ yields
\begin{eqnarray}
 &&\displaystyle\frac{|\det B|}{\pi^n }\int {d^n \xi\ }f(\xi)  \nonumber\\
& &  \times \int {d^n \mu}\, {\rm
e}^{i\left((\xi-\mu, B (\xi-\mu))-(q-\mu, B (q-\mu))\right)}
\nonumber\\
&=&\frac{|\det B|}{\pi^n }\int {d^n \xi\ }{d^n \mu}\,f(\xi) {\rm e}^{i\left[(\xi, B \xi)-(q, B q)+2\left(q-\xi, B \mu\right)\right]}
\nonumber\\
&=&\int {d^n \xi\ }f(\xi) {\rm e}^{i\left[(\xi, B \xi)-(q, B
q)\right]}\delta^n\left(q-\xi\right)=f(q).
\end{eqnarray}
The meaning of the above tomographic map depends on the features of
$B$. If we assume that $B$ is strictly positive (elliptic case),
this map corresponds to averages of $f$ along the ellipsoids defined
by Eq.\ (\ref{eq:quartic}). In particular if all the eigenvalues of
$B$ are equal to $b^2$ it corresponds to integration over spheres
centered at $\mu$ of (squared) radius $X/b^2$, namely,
\begin{equation}\label{eq:spheres}
b^2 (q-\mu)^2=X. \qquad (X >0)
\end{equation}

\begin{figure}[t]
\begin{center}
\includegraphics[width=\columnwidth]{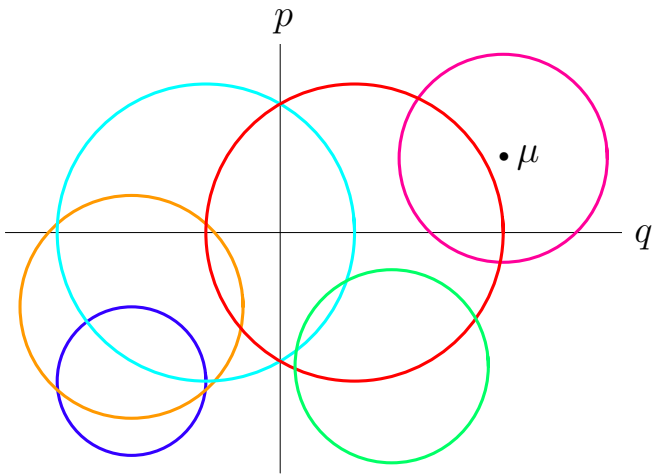}
\end{center}
\caption{Tomography on circles of center $\mu$ and (squared) radius $X/b^2$ on the plane.}
\label{fig:cylinder}
\end{figure}

Note that, in the two dimensional case, the distribution of circles
is different from that obtained by the transform defined by
diffeomorphisms in Sec.\ \ref{sec-submanifolds}. There, the family
of tomograms were defined only on circles passing through the
origin, including straight lines (circles of infinite radius). Here,
we are taking into account all possible circles of finite radius in
the plane (see Fig.\ \ref{fig:cylinder}). This corresponds to
trajectories of particles moving in a plane under the action of a
constant magnetic field. From a practical perspective, this new
tomographic map would make possible a different practical
implementation of tomography.

When $B$ has both positive and negative eigenvalues this correspond
to hyperbolic tomography with averages of $f$ along the hyperboloids
defined by Eq.\ (\ref{eq:quartic}), e.g.,
\begin{equation}\label{eq:hyperboloids}
b^2(q_1-\mu_1)^2-c^2(q_2-\mu_2)^2=X.
\end{equation}

In the case of degenerated $B$ forms we have to consider a hybrid transform.
$B$ can then be decomposed into a non-degenerated bilinear form and
a linear form. In this case the tomography of the linear components
should be treated as the standard Radon transform, whereas the
non-degenerate variables should  transform as above. Let us consider
for example a simple three-dimensional case with
\begin{equation}
\bar{B}=\left( \begin{array}{ccc} 1&0&0\\
0&1&0 \\ 0&0&0
\end{array}\right).
\end{equation}
In this case we can define the following tomographic map
\begin{eqnarray}
& &\omega_f(X,\mu;\bar{B}) = \int_{\mathbb{R}^3} {d^3 q}\, f(q) \nonumber\\
& &\quad \times \delta \left(X-(q_1-\mu_1)^2-(q_2-\mu_2)^2-\mu_3 q_3
\right),
\end{eqnarray}
with inverse transform
\begin{eqnarray}
f(q)&=&\int_{\mathbb{R}^4} \frac{dX\,d^3\mu}{2\pi^3}
\omega_f(X,\mu;\bar{B}) \nonumber\\
& & \quad \times {\rm e}^{i\left(X-(q_1-\mu_1)^2-(q_2-\mu_2)^2-\mu_3 q_3
\right)}.
\end{eqnarray}

\section{Conclusions and perspectives}
\label{sec-concl}

Let us discuss the main findings of this article, both from
mathematical and physical perspectives. From a mathematical
viewpoint, the generalizations of the Radon transform introduced
here enable one to consider marginals defined over submanifolds that
are not necessarily geodesic submanifolds in Riemannian spaces or
Lagrangian submanifolds of symplectic manifolds. These transforms
define tomograms over compact submanifolds and can be more suitable
for physical applications, because the practical implementation of a
tomogram can only achieved in local terms. In this new framework the
recovery of a local value (of a probability distribution on phase
space in the classical theory, or of a Wigner distribution in the
quantum case) only involves integration over manifolds that do not
reach infinity. In a previous article \cite{tomogram}, we considered
the tomography of a classical particle moving on a circle, which
required the definition of marginals over the helices on a cylinder.
Now, in the light of the new transforms just introduced, we have the
possibility of performing tomography over compact submanifolds even
for classical systems that evolve in unbounded domains. This is a
significant conceptual step forward.

Physically, the new reconstruction formulas enable one to generalize
the measurement procedures of the matter density of an object. In a
material medium with a strongly inhomogeneous refractive index, the
radiation beams (light beams, sonic beams or matter waves) would
propagate along curved lines and yet yield complete information on
the matter distribution by means of generalized Radon transforms.
For illustrative purposes our examples focus on two-dimensional
situations [see e.g.\ Eqs.\ (\ref{eq:conformal}),
(\ref{eq:inversion}) and (\ref{eq:hyperboloids})] but the approach
we propose is more general and valid in $\mathbb{R}^n$.

In quantum optics the new ``non-linear" Radon transforms can be
easily extended to the quantum domain by using the Weyl-Wigner map.
This will be discussed in a future publication. The results of this
article show that the reconstruction of the Wigner function using
optical or symplectic tomography based on straight-line Radon
transform can be extended to situations in which the marginals in
phase space are measured for curved hyperbolas or ellipses. In
particular, parabolic tomography could be implemented with the
recently observed accelerated Airy beams \cite{airy}.

Novel physical applications of tomography have attracted increasing
attention during the last few years. Recent applications involving
neutrinos, e.g.\ to get a mapping of the Earth inner density
\cite{kuo}, do not require new concepts of tomographic maps. However,
neutrino tomography of gamma ray bursts and massive stellar
collapses \cite{Meszaros} might require generalized tomography. In
particular, gamma rays tomography that made possible the discovery
of asphericity in supernovae explosions \cite{Maeda} or imaging of
astrophysical sources \cite{Zhang} can involve non-linear
trajectories of gamma rays due to strong gravitational lensing
effects. In those cases, generalizations of tomographic maps like
the ones considered in this paper are necessary.

\acknowledgments
We thank J.F.\ Cari\~nena and F.\ Ventriglia for helpful
discussions. V.I.M.\ was partially supported by Italian INFN and
thanks the Physics Department of the University of Naples for the
kind hospitality. P.F.\ and S.P.\ acknowledge the financial support
of the European Union through the Integrated Project EuroSQIP. The
work of M.A.\ and G.M.\ was partially supported by a cooperation
grant INFN-CICYT. M.A. was also partially supported by the Spanish
CICYT grant FPA2006-2315  and  DGIID-DGA (grant2006-E24/2). The work
of V.I.M. was partially supported by the Russian Foundation for
Basic Research under Project No.\ 07-02-00598.


\end{document}